\documentclass{appolb} \usepackage{epsfig}
\newcommand{\order}[1]{\mathcal O\left(#1\right)}
\newcommand{\BXsg}{\overline B \to X_s \gamma}
\newcommand{\bsg}{b \to s \gamma}
\newcommand{\nn}{\nonumber \\}

\newcommand{\ep}{\varepsilon}

\newcommand{\scs}{\scriptscriptstyle}

\newcommand{\Leff}{{\mathcal L}_{\rm eff}}
\newcommand{\LQCDQED}{{\mathcal L}_{{\rm QCD} \times {\rm QED}} 
 (u, d, s,c, b)}

\begin{document}
% \eqsec  % uncomment this line to get equations numbered by (sec.num)
\title{Progress in the evaluation of the $\BXsg$ decay 
rate at NNLO%
\thanks{Presented at XXXI Conference of Theoretical Physics: "Matter
To The Deepest", Ustro\'n, Poland, 5.-11. September 2007}%
~\thanks{Work supported by the Sofia Kovalevskaja programme of the
Alexander van Humboldt Foundation sponsored by the German Federal
Ministry of Education and Research}
}
\author{Thomas Schutzmeier
\address{Institut f\"ur Theoretische Physik und Astrophysik,
Universit\"at W\"urzburg\\
Am Hubland, D-97074 W\"urzburg, Germany}
}
\maketitle

\begin{abstract}
The theoretical status of NNLO QCD corrections to 
the inclusive radiative $\BXsg$ decay in the standard model is 
briefly overviewed. Emphasis is put on recent results for three-loop 
fermionic corrections to matrix elements of the most
relevant four-quark operators.
\end{abstract}
  
\section{Introduction}
The inclusive $\BXsg$ decay mode, a flavor-changing-neutral-current
process and therefore loop-supressed in the standard model (SM),
is known to be a sensitive probe of new physics.
Obviously, deriving constraints on the parameter space of physics beyond the 
SM relies strongly on both accurate measurements and precise 
theory predicitions within the SM. 

Combining measurements of BaBar, Belle and CLEO \cite{ref:exp},
the current world average for the branching ratio with a
cut $E_{\gamma,0} > 1.6\,\mbox{GeV}$ on the photon energy in 
the $\overline B$-meson rest frame reads \cite{Group:2007cr}
\begin{equation}\label{eq:HFAG}
 \mathcal B(\bar B \to X_s \gamma)_{\scs E_{\gamma} > 1.6\,{\rm
 GeV}}^{\scs\rm exp } = \left(3.55\pm
 0.24{\;}^{+0.09}_{-0.10}\pm0.03\right)\times 10^{-4},
\end{equation}
where the first uncertainty corresponds to a combined statistical and
systematic error, the second one is due to the theory input in the
extrapolation of the measured branching ratio to the reference value
$E_{\gamma,0}$, whereas the third one is connected to the subtraction
of $b\to d\gamma$ events. The overall error of the world average
amounts to about 7\% which is comparable with the expected size of
next-to-next-to-leading order (NNLO) QCD effects to the perturbative
transition $b \to X_s^{parton} \gamma$. Thus, a complete
SM calculation at this accuracy level is desired.

To a large extent, the NNLO program has been finished and the latest
theoretical estimate based on the results \cite{Misiak:2006zs}
\begin{equation}\label{theoretical B}
    {\mathcal B}({\bar B}\to X_s\gamma)_{\scs E_{\gamma} > 1.6\,{\rm GeV}}^{\scs\rm theo} = (3.15 \pm 0.23) \times 10^{-4}
\end{equation}
is in good agreement with the experimental value eq.~(\ref{eq:HFAG}).
Here, the error consists of four types of uncertainties added in
quadrature: non-perturbative (5\%), paramteric (3\%), higher-order
(3\%) and $m_c$-interpolation ambiguity~(3\%). 

\section{The effective theory framework}

The partonic decay width $\Gamma(\bsg)$
recieves large contributions of logarithms $\log M_W^2/m_b^2$.
Resumming them at each order of $\alpha_s$ is most
suitably done in the framework of an effective low-energy theory with
five active quarks by integrationg out the top and heavy electroweak fields. 
The relevant effective Lagrangian is given
by 
\begin{equation}\label{eq:effectivelagrangian}
\Leff = \LQCDQED + \frac{4G_F}{\sqrt{2}} V^\ast_{ts} V_{tb} 
\sum^{8}_{i= 1} C_i (\mu) \, Q_i (\mu).
\end{equation}
The usual QED-QCD Lagrangian for the light SM fields is stated in the
first term whereas the second term gives the local operator product expansion
(OPE) with Wilson coefficients $C_i(\mu)$ and operators $Q_i(\mu)$ up to 
dimension six built out of the light fields. $V_{ij}$ denotes elements of the
Cabibbo-Kobayashi-Maskawa matrix and  $G_F$ the Fermi coupling constant.

The operator basis reads
\begin{eqnarray} \label{eq:physicaloperators}
Q_{1,2} \, & = & \, (\bar{s} \Gamma_i c) (\bar{c} \Gamma'_i b), \nn 
Q_{3,4,5,6} \, & = & \, (\bar{s} \Gamma_i b) \sum\nolimits_q (\bar{q} \Gamma'_i
q) ,\nn 
Q_7 \, & = & \, \frac{e}{16 \pi^2}\,{\overline m_b}(\mu)\, (\bar{s}_L 
  \sigma^{\mu \nu} b_R) \, F_{\mu \nu} , \nn
Q_8 \, & = & \, \frac{g}{16 \pi^2}\,{\overline m_b}(\mu)\, (\bar{s}_L 
  \sigma^{\mu \nu}T^a b_R) \, G^a_{\mu \nu}.
\end{eqnarray}
where $\Gamma$ and $\Gamma'$ represent various products of Dirac and 
color matrices. $\overline m_b(\mu)$ is the bottom mass in the 
$\overline{MS}$ scheme and the sum runs over all light quark
flavours $q$.  

Consistent calculations of $\Gamma(\bsg)$ in the effective framework are 
performed in three steps.
Teh Wilson coefficients $C_i(\mu_0)$
$\mu_0 \approx M_W$ are first determined  at the electroweak scale by
requiring equality of Green's  
functions in the effective and full theory at leading order in
(external~momenta)/$M_W$. Subsequently, the operator mixing under
renormalization is computed by evolving $C_i(\mu)$ from $\mu_0$ down
to the low-energy scale $\mu_b$ with help of effective theory
Renormalization Group Equations (RGE). Finally, the matrix elements with
single  
insertions of effective operators are computed.
Non-perturbative effects appear only as small corrections to the last
step, which is connected to the heaviness of the bottom quark and the
inclusiveness of the $\BXsg$ decay mode.\\
As far as the next-to-leading order precision is concerned,  
this program has been completed already a few years ago, 
thanks to the joint effort of many groups 
(see for eg.~\cite{Buras:2002er,Hurth:2003vb} and references 
therein). 
The NNLO calculation, which involves 
hundreds of three-loop on-shell vertex-diagrams and thousands of four-loop
tadpole-diagrams, is a very complicated task and, as already mentioned
in the introduction, large parts have already been finished.\\
Matching the four-quark operators $Q_1,...,Q_6$ and the dipole
operators $Q_7$ and $Q_8$ at the two- and three-loop level,
respectively, has been performed in \cite{Bobeth:1999mk,Misiak:2004ew}. 
The three-loop renormalization in the $\{Q_1, \dots ,Q_6\}$ and
$\{Q_7, Q_8 \}$ sectors was found
in~\cite{Gorbahn:2004my,Gorbahn:2005sa}, and results for the four-loop
mixing of $Q_1, \dots, Q_6$ into $Q_7$ and $Q_8$ were lately
provided in~\cite{Czakon:2006ss} completing the anomalous-dimension
matrix.
The two-loop matrix element of the photonic
dipole operator $Q_7$ was found, together with the corresponding 
bremsstrahlung, in~\cite{Melnikov:2005bx,Blokland:2005uk}
and confirmed in~\cite{Asatrian:2006ph}.
Moreover, contributions of the dominant operators in the so-called
large-$\beta_0$ approximation ($\order{\alpha_s^2 \beta_0}$)
to the photon energy spectrum have been computed in~\cite{Ligeti:1999ea}.
Three-loop matrix elements of the operators $Q_1$ and $Q_2$ at
$\order{\alpha_s^2 \beta_0}$ and two-loop matrix elements of
$Q_7$ and $Q_8$ were found in~\cite{Bieri:2003ue} as expansions in the
quark mass ratio $m_c^2/m_b^2$. Recently, we confirmed
the findings of~\cite{Bieri:2003ue} on the matrix
elements of $Q_{1,2}$ and were able to extend the calculation 
beyond the large-$\beta_0$ approximation by evaluating the 
full fermionic contributions \cite{Boughezal:2007ny}. This calculation is briefly reviewed below.
Furthermore, in \cite{Misiak:2006ab}, the full matrix elements
of $Q_1$ and $Q_2$ have been computed in the large $m_c$ limit, 
$m_c \gg m_b$, and subsequently used to perform an interpolation 
to the physical region assuming that the large-$\beta_0$ part is a good
approximation at $m_c=0$. This is the source of the interpolation
ambiguity mentioned beneath eq.~(\ref{theoretical B}).

\section{NNLO fermionic corrections to the matrix elements of $Q_{1,2}$}
Matrix elements of $Q_1$ and $Q_2$ constitute a crucial input for the
accuracy of the current NNLO estimate eq.~(\ref{theoretical B}). The
intention of our recent work~\cite{Boughezal:2007ny} was the determination of full fermionic corrections to these matrix elements to cross-check the results of~\cite{Bieri:2003ue}
and, at the same time, to test the validity of the massless
approximation used in the large-$\beta_0$ approximation. 
\begin{figure}[t]
 \begin{center}
 \begin{minipage}{.45\textwidth}
  \epsfig{file=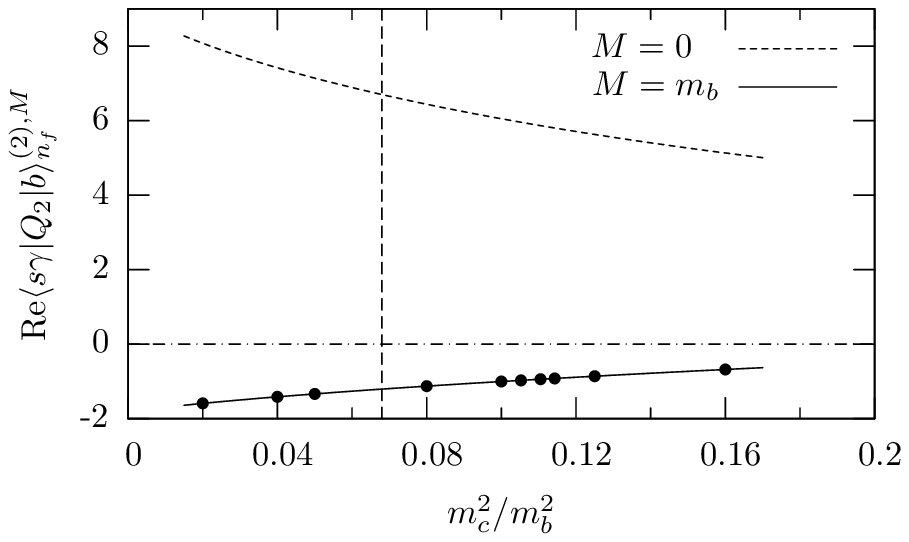, width=1.\textwidth}
 \end{minipage}
 \begin{minipage}{.45\textwidth}
  \epsfig{file=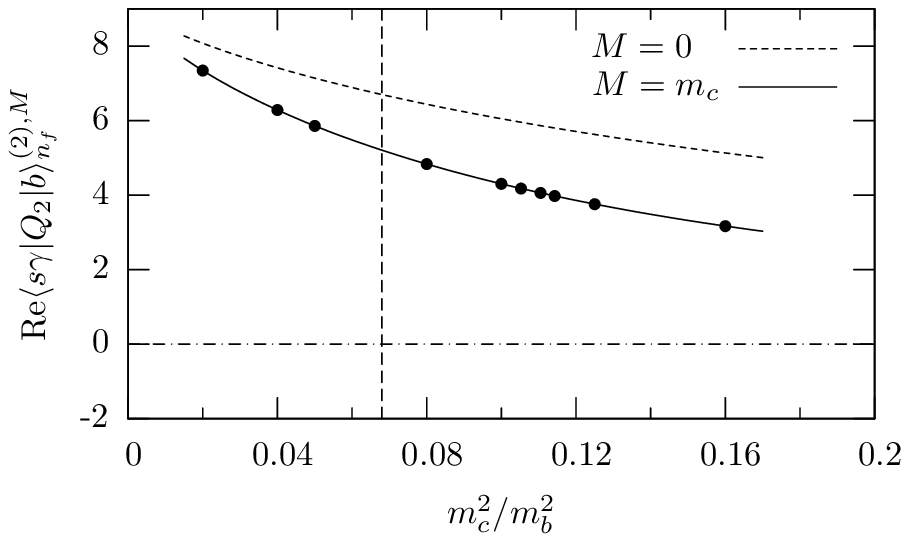, width=1.\textwidth}
 \end{minipage}
 \end{center}
 \caption{Plots of the $\mathcal O(\alpha_s^2 n_f)$ corrections to
 matrix elements of $Q_1$ as function of $z=m_c^2/m_b^2$ with fermionic
 loops of mass $M$ and $\mu=m_b$. (a):~$M=m_b$, (b):~$M=m_c$. 
 The $M=0$ case is also shown for comparison. \label{fig:plots}}
\end{figure}
Our calculcation is based on two different techniques, that were
applied to the master integrals obtained from IBP reduction.
In the case of massless quark loop insertions into the gluon
propagator of the relevant NLO diagrams, 
all integrals have been performed using the
Mellin-Barnes (MB) method. The MB representations were derived using
an automated package \cite{chachamis-czakon:2006} and 
analytically continued with help of the \texttt{MB} package
\cite{Czakon:2005rk}.
After expanding in $z=m_c^2/m_b^2$ the resulting coefficients represented as
series over residues could be resummed with \texttt{XSummer}
\cite{Moch:2005uc}.
In addition, an exact solution through direct numerical integration keeping the
full $z$-dependence was obtained.
This procedure was not applicable in the case of massive quark loop
insertions due to poor convergence. Instead, the method of differential 
equations as a second approach was utilized. 
Using the fact that the master integrals $V_i(z,\epsilon)$ (after
rescaling by a trivial factor) are
functions of $\epsilon$ and the mass ratio $z^{-1}$ a system of differential equations has been generated where the right-hand side was again expressed through master integrals with the help of relations obtained from the reduction. 
The solution of this system for 
arbitrary values of $y$ proceeded in two steps. First, an expansion 
in $\ep$ and $y$ for $\epsilon,\, y\rightarrow 0$ was performed 
and the coefficients were calculated recursively up to high
powers of $y$~\cite{Boughezal:2006uu}. Using the resulting high
precision values at a starting point $y \ll 1$, the unexpanded system
was integrated numerically up to physical values of $y$ with help of
the Fortran package ODEPACK \cite{odepack}. The path was shifted into
the complex plane to avoid special points.  
Figure \ref{fig:plots} shows the resulting data points together with
fits for the renormalized matrix elements of $Q_1$ with an internal
quark of mass $M=0,m_c,m_b$. In the case $M=m_b$ it is evident that the
massless approximation overestimates the massive result by a large factor. 
For $M=m_c$ this difference is not that pronounced but still not negligible.

\section{Conclusions}
Taking new results for the full fermionic corrections at NNLO 
into account, the branching ratio is enhanced by about one 
to two percent in comparison to \cite{Misiak:2006ab}. 
Moreover, an evaluation of bosonic corrections at this order, thereby 
completing three-loop matrix elements, is essential to resolve the 
interpolation ambiguity and to further improve the SM prediction for
the $\BXsg$ decay.

\end{document}